\documentclass[fleqn,usenatbib]{mnras}

\usepackage{newtxtext,newtxmath}
\usepackage[T1]{fontenc}
\DeclareRobustCommand{\VAN}[3]{#2}
\let\VANthebibliography\thebibliography
\def\thebibliography{\DeclareRobustCommand{\VAN}[3]{##3}\VANthebibliography}

\usepackage{graphicx}	
\usepackage{amsmath}	

\title[
IXPE view of LMC~X$-$3]{Unveiling the X-ray polarimetric properties of LMC~X$-$3 with \textit{IXPE}, \textit{NICER}, and \textit{Swift}/XRT}
\author[Garg, Rawat, $\&$ M\'endez]{Akash Garg$^{1}$\thanks{E-mail: akash.garg@iucaa.in (AG)},
Divya Rawat$^{1,2}$,
Mariano M\'endez$^{3}$,
\\
$^{1}$Inter-University Center for Astronomy and Astrophysics, Ganeshkhind, Pune 411007, India\\
$^{2}$Observatoire Astronomique de Strasbourg, Universit\'e de Strasbourg,
CNRS, 11 rue de l’Universit\'e, F-67000 Strasbourg, France\\
$^{3}$Kapteyn Astronomical Institute, University of Groningen, PO BOX 800, Groningen NL-9700 AV, the Netherlands
}

\date{Accepted XXX. Received YYY; in original form ZZZ}

\pubyear{2023}

\begin{document}
\label{firstpage}
\pagerange{\pageref{firstpage}--\pageref{lastpage}}
\maketitle
\begin{abstract}
The incoming Imaging X-ray Polarimetry Explorer (IXPE) observations of X-ray binaries provide a new tool to investigate the underlying accretion geometry. Here we report the first measurements of X-ray polarization of the extra-galactic black-hole X-ray binary LMC X$-$3. We find a polarization fraction of $\sim 3 \%$ at a polarization angle of $\sim 135^\circ$ in the $2-8$ keV energy band with statistical significance at the 7$\sigma$ level. This polarization measurement significantly exceeds the minimum detectable polarization threshold of 1.2\% for the source, ascertained at a 99\% confidence level within the $2-8$ keV energy band. The simultaneous spectro-polarimetric fitting of \textit{NICER}, \textit{Swift}/XRT, and \textit{IXPE} revealed the presence of a disc with a temperature of $\sim$1 keV and a Comptonized component with a power-law index of $\sim$ 2.4, confirming the soft nature of the source. The polarization degree increases with energy from $\sim$3\% in the $2-5.7$ keV band to $\sim$9\% in the $5.7-8$ keV band, while the polarization angle is energy independent. The observed energy dependence and the sudden jump of polarization fraction above 5 keV supports the idea of a static slab coronal geometry for the Comptonizing medium of LMC X$-$3.
\end{abstract}
\begin{keywords}
accretion: accretion discs; polarization; X-rays: binaries; X-rays: individual: LMC~X$-$3
\end{keywords}



\section{Introduction}
Black hole X-ray binaries (BXB) are an exquisite class of astronomical systems in which a black hole accretes matter from a companion star and forms a planar structure known as an accretion disc. Such an accretion disc is a powerful source of electromagnetic radiation, emitting profusely in X-ray, optical and radio wavelengths. During its outburst, the BXB transitions from a low-hard state (LHS, low flux of high energy photons) to a high soft state (HSS, high flux of low energy photons) via hard/soft intermediates states and sometimes, steep-power law (SPL) state \citep[see][]{re06,mo12}. These states are characterised by varying proportions of count rate and flux hardness \citep[][and references within]{be05,be11}. In the HSS, the spectrum is primarily characterized by an optically thick thermal component, often modelled as a multi-temperature blackbody of temperature $\sim$ 1 keV \citep{shakura_1973}. Occasionally, a power law component with a power law index $\Gamma \ge$2  is also observed alongside the thermal component \citep{me97,do07}. In contrast, during the LHS, the X-ray spectrum exhibits a distinct feature attributed to Comptonization from an electron plasma with temperatures between $50-100$ keV  \citep{gi10}.

Over more than five decades, the X-ray spectro-timing analysis has proven to be an important tool to map the accretion region indirectly. Besides establishing the presence of a cold accretion disc for the soft component, different models such as the slab-corona model, patchy corona model, magnetized accretion ejection model, and more exist to justify the hard component in the spectra \citep[][and references within]{haardt_1993,do07}. However, it is important to note that the inherent ambiguity regarding the nature of the Comptonizing medium cannot be fully resolved through X-ray spectro-timing analysis alone. Given this scenario, the X-ray polarization study 
 offers an additional tool to break the existing degeneracies among the predictions for the accretion flow around the black hole. The linear polarization gives two extra parameters, polarization angle and polarization degree, whose variations with energy and time can be predicted and checked against the observations.
 
  Over the last few decades, numerous models have been proposed such as Monte-Carlo polarization simulations for a thermal accretion disc \citep{sc09} and a wedge, clumpy or spherical corona \citep{Schnittman_2010}.  \citet{Krawczynski_Beheshtipour_2022} employed the {\sc{KerrC}} model to determine polarization fractions within geometrically thin discs and examined the impact of varying coronal shapes. \citet{poutanen_2023} applied radiative transfer simulations on the polarization spectrum for static and dynamical coronal regions. \citet{po22} explored the effects of relativistic polarization using the {\sc{kynstokes}} code for lamp-post corona model. Furthermore, \citet{zhang19} employed the \textsc{Monk} code to investigate Comptonization within the framework of Kerr spacetime. Besides these, there are a few more studies that probe the accretion geometry in BXB using spectro-polarimetric signals. The launch of the \textit{IXPE} (Imaging X-ray Polarimetry Explorer) mission provides the opportunity to test these models. Therefore, a comprehensive polarimetric study is necessary to untangle the geometry of the accretion and Comptonizing medium within these systems.

The recent \textit{IXPE} 
observations of the BXB 4U 1630$-$47 have provided intriguing results. A polarization fraction of $\sim 8\%$ in the HSS \citep{rawat_2023,Kushwaha_2023,Ratheesh_2023} and $\sim 7\%$ in the SPL state \citep{rawat_2023b,rodriguez_2023} was observed. This observed high polarization fraction in the HSS has captured the attention of X-ray astrophysicists, as it defies explanations within the framework of contemporary models \citep{Schnittman_2010,ta20, Krawczynski_Beheshtipour_2022}. Similarly, using \textit{IXPE}, \citet{kr22}  has reported a polarization fraction of $4\%$ in the hard state of Cygnus X$-$1, which is unusual for a low inclination source assuming polarization is due to electron scattering. 

Motivated by the recently reported high polarization fraction for BXB in the high soft state, we explore the polarimetric properties of LMC~X$-$3 in this paper. LMC~X$-$3 is a black hole X-ray binary source discovered in the Large Magellanic Cloud with Uhuru \citep{leong_1971}. The binary system has an inclination angle of $\sim 70^\circ$ \citep{orosz_2014}. After its discovery, the spectral analysis by \citet{white_1984} showed that LMC~X$-$3 has an unusual soft X-ray spectrum, implying that the source is in the soft state for most of the time, comprising of an ultrasoft and a high-energy component. Later, LMC X-3 was also observed to undergo brief low/hard states \citep[LHS, defined by simple power-law,][]{boyd20,wilm01,bhuv22} and anomalous low states \citep[ALSs,][]{smale12,torpin17}. \citet{fender_2000} conducted a radio survey on persistent BXB sources and reported no radio counterpart of LMC~X$-$3, which could be attributed to either the distant location of the source ($\sim$48 kpc, \citealt{orosz_2014}) or its unusually soft spectrum. The dimensionless spin parameter of the black hole in this source, using different missions, is reported to be in the range $\sim 0.19-0.24$ \citep{Steiner_2014,bhuv22,jana21}.\\

For highly inclined sources, if the polarization is due to electron scattering, then a high polarization fraction is expected with the polarization angle aligned parallel to the disk plane \citep{ch47,ch60,so63}. The high inclination angle and the fact that LMC~X$-$3 spends most of its time in the soft state make it an excellent candidate for a polarimetric study. The subsequent sections of this paper are organized as follows: Section~\ref{obs} covers the observations and data reduction methods, Section~\ref{results} presents polarimetric and spectro-polarimetric findings, and in Section~\ref{discuss} we conclude with interpretations and implications of our results for LMC X-3.

\begin{figure}
\centering\includegraphics[scale=0.37,angle=0]{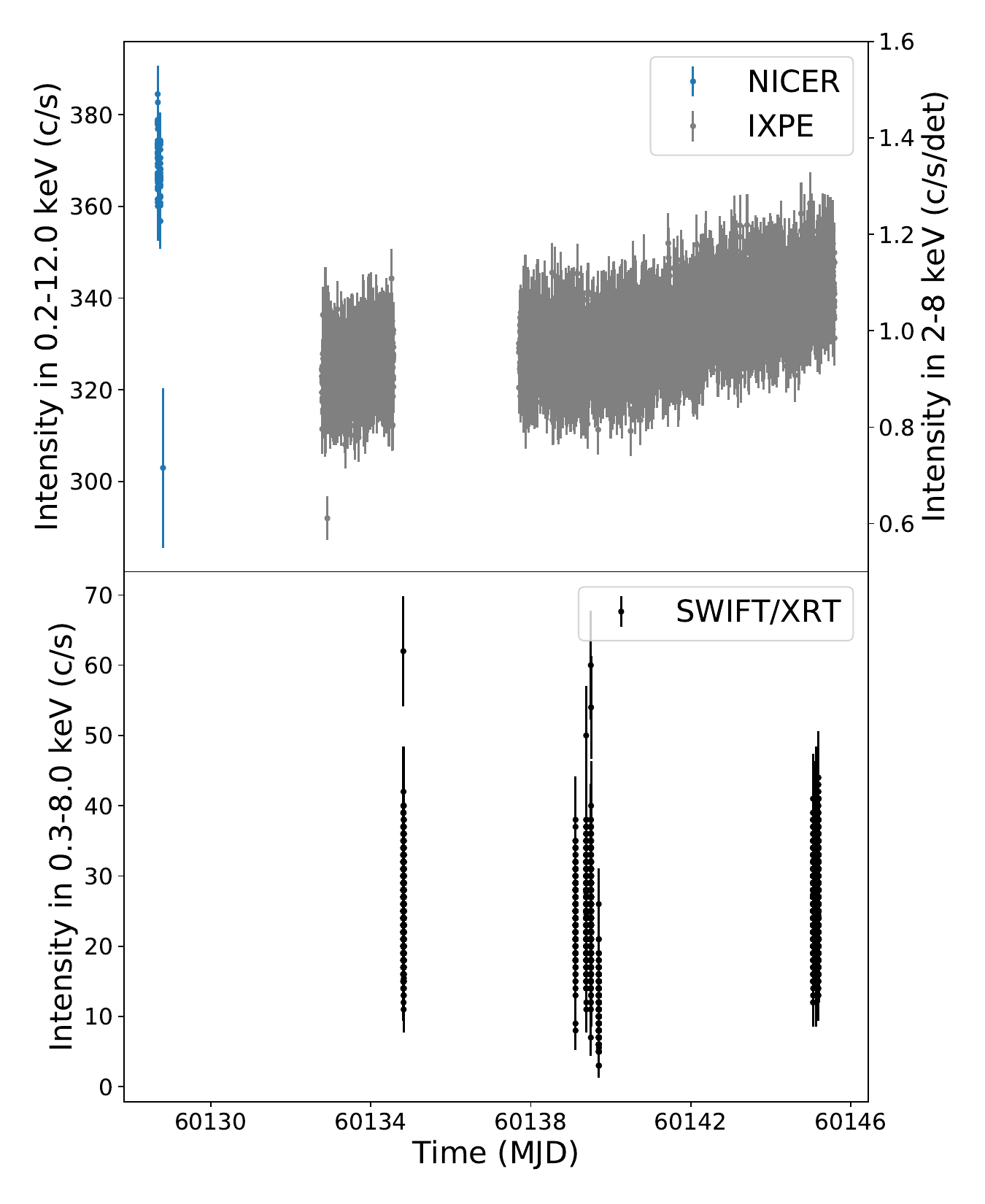}
\caption{Top and bottom panels show the {\it{NICER}} (10 s bin), {\it{IXPE}} (100 s bin), and {\it{Swift/XRT}} (1 s bin) light curves of LMC~X$-$3 in the $0.2-12$~keV, $2-8$~keV and $0.3-8$~keV bands, respectively.}
\label{lightcurve}
\end{figure}
\begin{figure}
\centering\includegraphics[scale=0.34,angle=0]{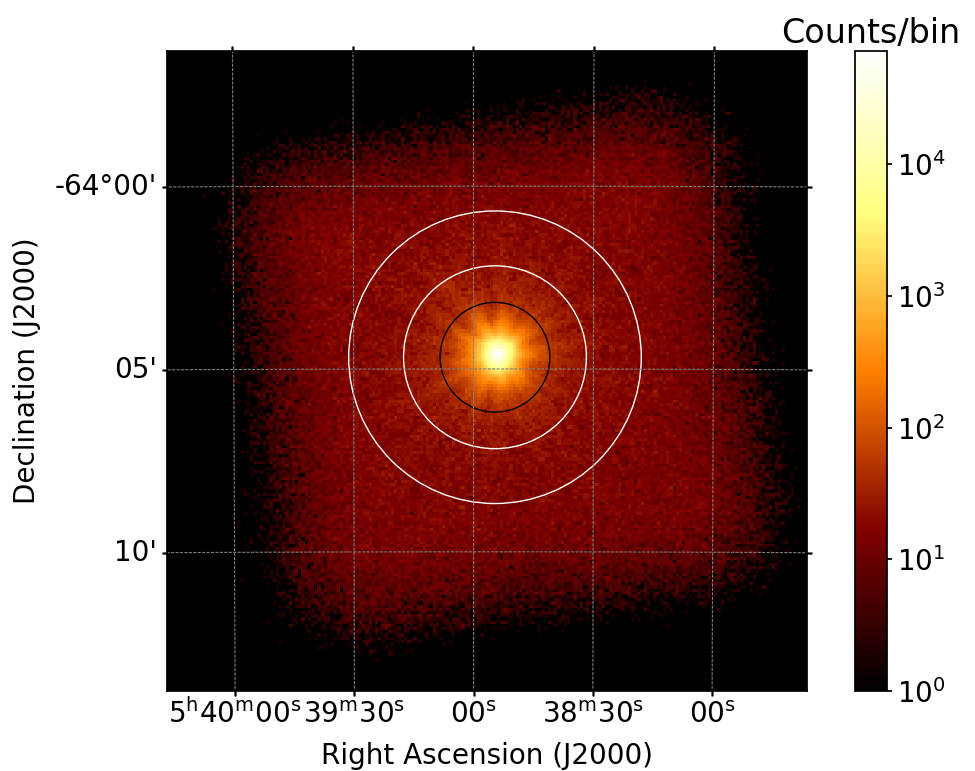}
\caption{\textit{IXPE} count map of detector DU1. The black and white circles on the image show the circular source and annular background regions of LMC X-3.}
\label{count_image}
\end{figure}

\section{Observation and  Data Reduction}
\label{obs}
In this work, we have used \textit{NICER}, \textit{Swift}/XRT and \textit{IXPE} observations of the source LMC~X$-$3 taken during July 2023. The observation details with each instrument are given in Table \ref{obs_log}, and the data reduction techniques are briefly discussed in the next subsections.

\subsection{NICER}
We have analyzed observations of LMC~X$-$3 with the Neutron Star Interior Composition Explorer \citep[\textit{NICER}][]{ge12} on 2023-07-03. \textit{NICER}'s XTI \citep[X-ray Timing Instrument][]{ge16} covers the $0.2-12.0$ keV band and has an effective area of $>$2000 cm$^{2}$ at 1.5 keV. The energy and time resolutions are 85 eV at 1 keV and 4 $\times 10^{-8}$ s, respectively. We have applied the standard calibration process and screening using the {\sc{nicerl2}}\footnote{\url{https://heasarc.gsfc.nasa.gov/docs/nicer/analysis_threads/nicerl2/}} task. Further, the grouped spectrum is extracted using {\sc{nicerl3}}\footnote{\url{https://heasarc.gsfc.nasa.gov/docs/nicer/analysis_threads/nicerl3-spect/}}, which also applies systematic errors and quality flags to the spectrum. In addition, {\sc{nicerl3}} also creates ancillary (ARF), response (RMF), and background files (for the 3C50 model) using CALDB version 20221001\footnote{\url{https://heasarc.gsfc.nasa.gov/docs/heasarc/caldb/nicer/}}.

\subsection{Swift/XRT}
\textit{Swift}/XRT has observed the LMC~X$-$3 in the window mode for three epochs between 2023-07-09 and 2023-07-20. \textit{Swift}/XRT \citep{burrow_2005} is an X-ray Imaging Telescope that operates in a narrow energy band of 0.2–10~keV with an effective area of $\sim$125 cm$^2$ at 1.5~keV. We have used {\sc{xrtpipeline}} to extract the clean event files which are then used to extract the lightcurve using the {\sc{xselect}} (V2.5b) package of {\sc{heasoft}} version 6.32.1\footnote{\url{https://heasarc.gsfc.nasa.gov/lheasoft/download.html}}. Utilizing an online \textit{Swift}/XRT tool {\sc{Build Swift-XRT products}}\footnote{\url{https://www.swift.ac.uk/user_objects/}}, we have created a single averaged spectrum from the three epochs along with ancillary response (arf) and response (rmf; see \citealt{ev09}) files. We have considered grade $0-2$ events for the source spectra as the WT mode data should not be affected by pile-up for source count below 100 c/s \citep{romano2006}.

\begin{table}

 \caption{Observation log for LMC~X$-$3. The columns are the Instruments used, their ObsID, the observation's start and end date, and the exposure times.}
 \begin{center}
\scalebox{0.95}{%
\begin{tabular}{ccccc}
\hline
Instrument &ObsID & Tstart & Tstop & exposure \\
& & (MJD) & (MJD) & (secs)\\ 
\hline

NICER & 6101010115 & 60128.67 & 60128.81 & 630 \\
\hline

Swift/XRT & 00089714001 & 60134.81  & 60134.83 & 1244 \\
& 00089714002 & 60139.12 & 60139.70  & 1772  \\
& 00089714003 & 60145.06 & 60145.20 &  1928 \\

\hline
IXPE & 02006599 & 60132.78 & 60145.60 & 561925
 \\
\hline
\end{tabular}}
\end{center}
\label{obs_log}
\end{table}

 \begin{figure}
\centering\includegraphics[scale=0.35,angle=0]{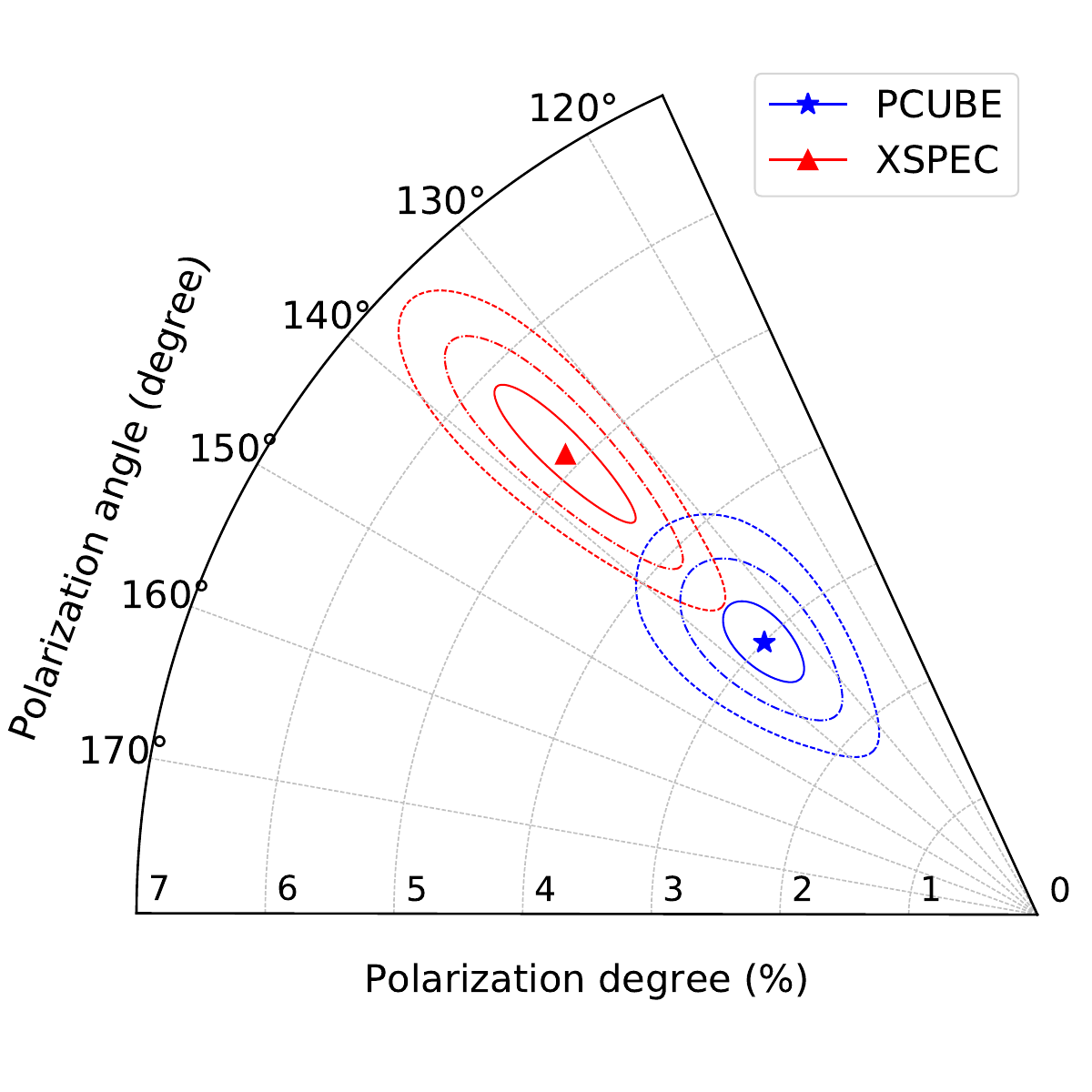}
\caption{Contour plots of PA and PD of LMC~X$-$3 in the 2--8 keV band with events from all DUs summed up using the {\sc{pcube}} algorithm, and the {\sc{xspec}} fitting method. The 3 contours represent confidence levels of 68.27\%, 95.45\%, and 99.73\%.}
\label{2_8_stoke}
\end{figure}
%
 \begin{figure}
\centering\includegraphics[scale=0.34,angle=0]{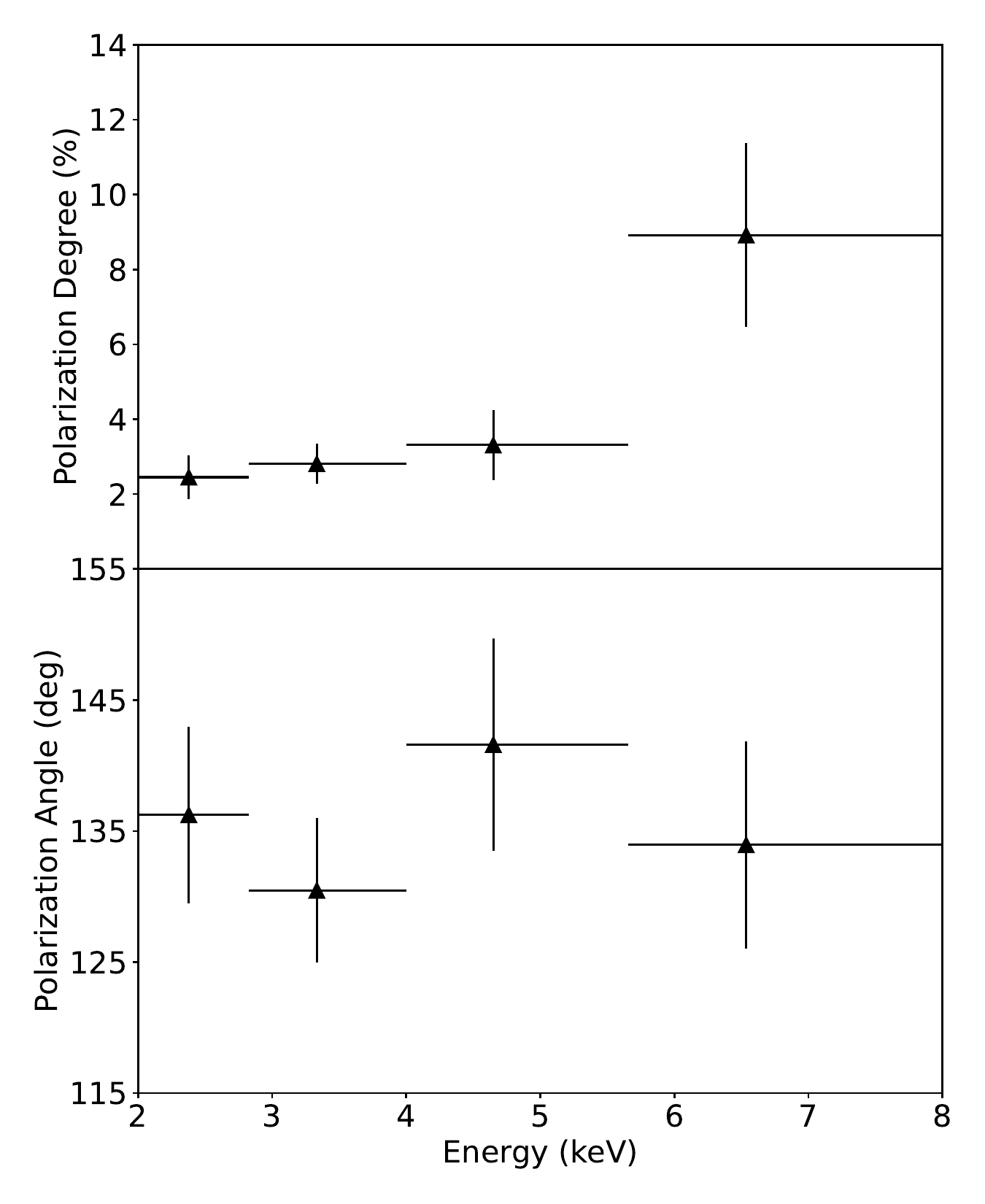}
\caption{The PD (Top Panel) and PA (Bottom panel) of LMC~X$-$3 as a function of energy with events from all DUs summed up using the {\sc{pcube}} algorithm.}
\label{pa_pd_energy}
\end{figure}

 \begin{figure*}
\centering\includegraphics[scale=0.29,angle=0]{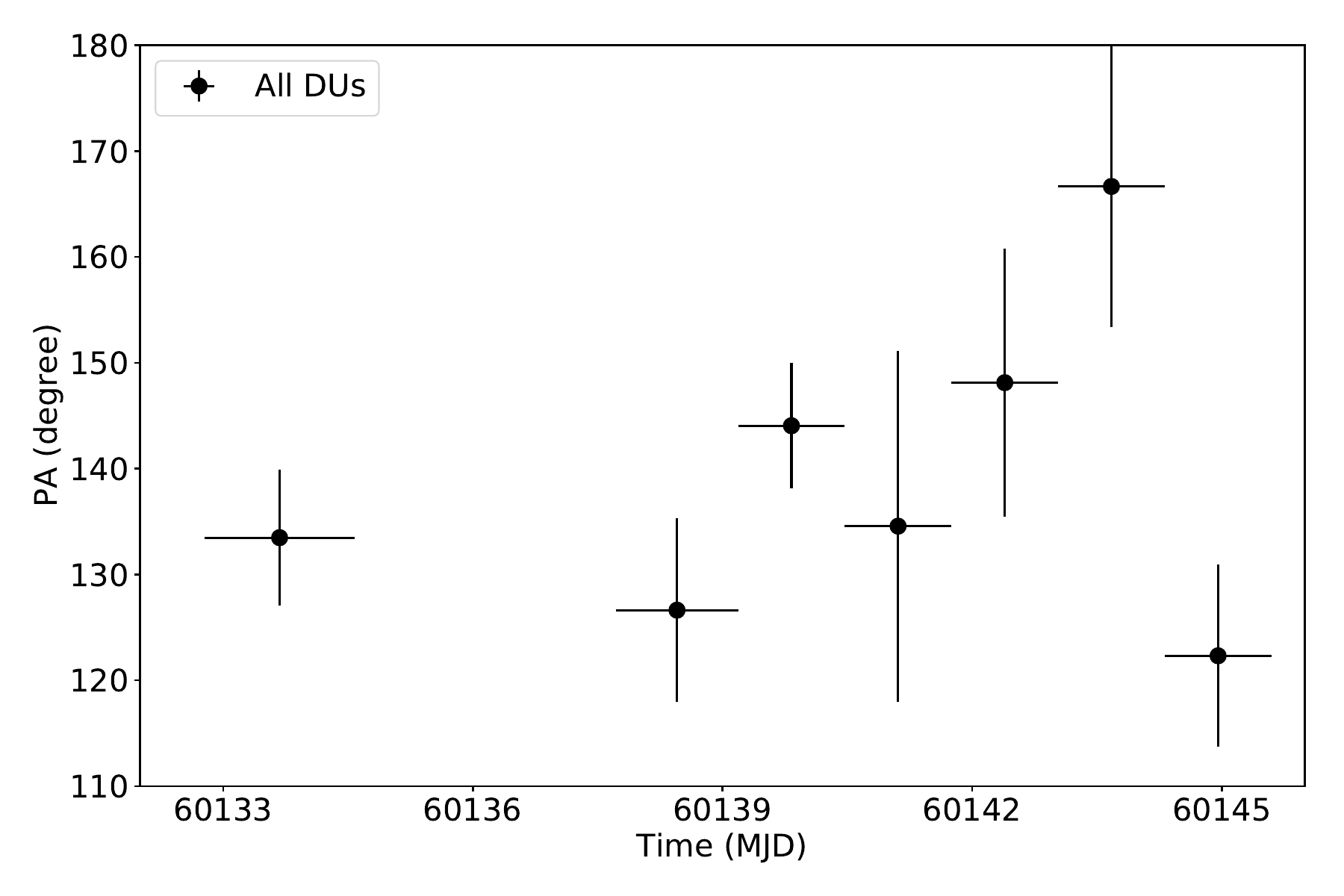}
\centering\includegraphics[scale=0.29,angle=0]{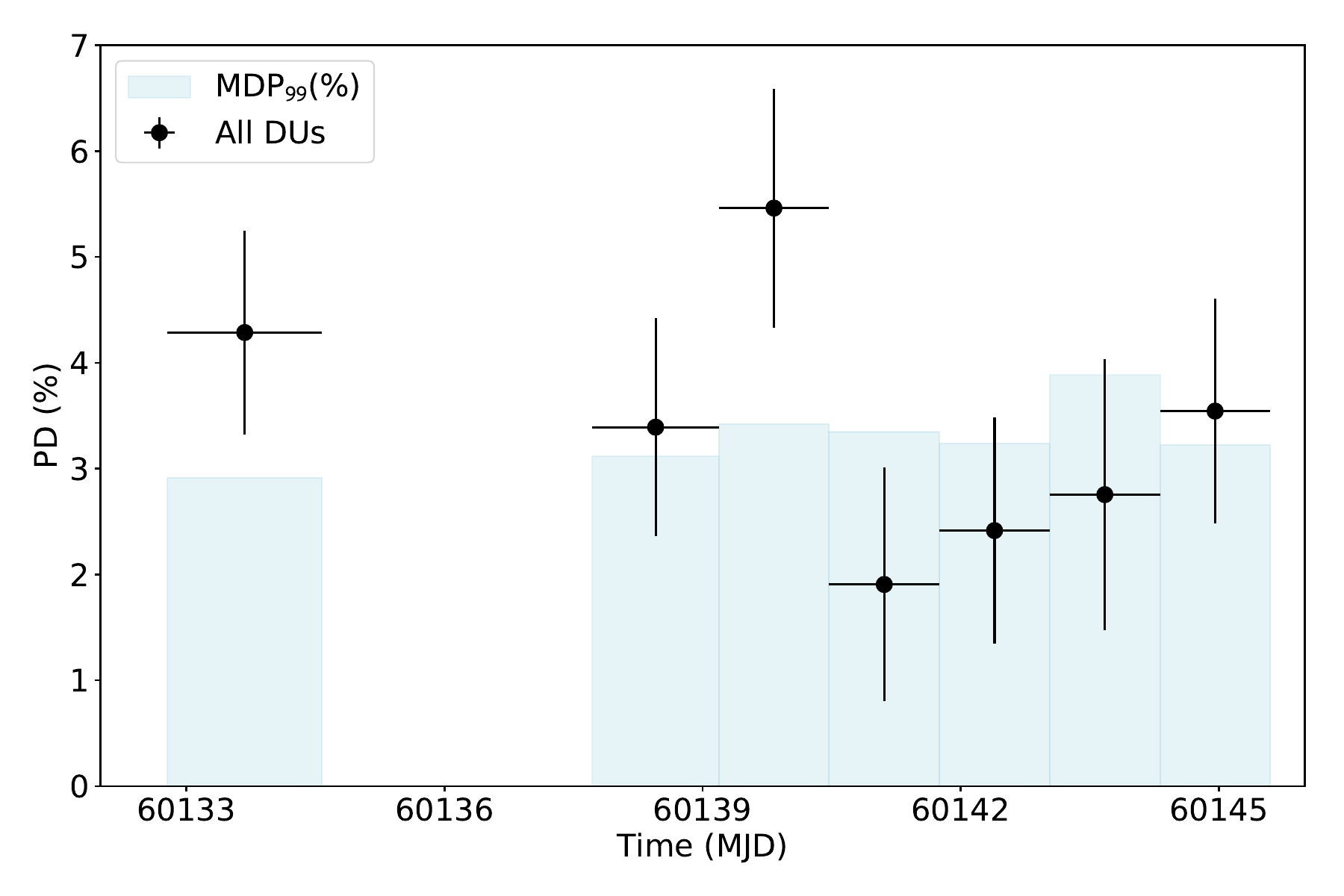}
\caption{The PA (Left panel) and PD (Right Panel) of LMC~X$-$3 as a function of time with events from all DUs summed up using the {\sc{pcube}} algorithm. The lightblue shaded region in the Right panel represents MDP$_{99}$ for each time bin.}
\label{pa_pd_time}
\end{figure*}

\subsection{IXPE}
\textit{IXPE} is a recently launched X-ray polarimetry observatory developed by NASA in collaboration with the Italian Space Agency (ASI). It can measure the polarization properties of multiple X-ray sources in the energy band of $2-8$~keV. Besides polarization, \textit{IXPE} also
captures the spatial, temporal and spectral aspects of X-ray photons \citep{ma20,ba21,di22,we22}. \textit{IXPE} has observed the source LMC~X$-$3 from 2023-07-07 to 2023-07-20 for a total exposure of $\sim$ 562 ksec. Figure-\ref{lightcurve} shows the \textit{NICER} (top panel) and \textit{Swift}/XRT (bottom panel) lightcurves in the $0.2-12$ keV and $0.3-8$ keV bands, respectively along with the simultaneous $2-8$ keV averaged \textit{IXPE} lightcurve (top panel) extracted for the source region from all three detectors. We note here that the \textit{NICER} observation is not simultaneous with \textit{IXPE} but we still consider it for spectro-polarimetric analysis owing to minimal changes in source flux and spectral shape during the observation period. 

We used the \textit{IXPE} level-2 event files for all the detector units (DU) obtained directly from the HEASARC archive. We begin by extracting and plotting the \textit{IXPE} count image of event files using {\sc{xselect}} and {\sc{ds9}}, respectively, to choose the appropriate source and background regions as we show in Figure-\ref{count_image} \citep[for more details, see][]{Podgorny_2023_LMX_X1}. Further, we employ these regions (a circle of radius 1.5 arcmin for source and an annulus of inner and outer radii of 2.5 arcmins and 4 arcmins, respectively, for the background) in the {\sc{xpselect}} tool of {\sc{ixpeobssim}} 30.2.2\footnote{\url{https://ixpeobssim.readthedocs.io/en/latest/overview.html}} \citep{ba22} to produce the cleaned source and background event files for all DUs. Then we bin the event files using different algorithms of {\sc{xpbin}} \citep{kis15} to extract the polarization properties of the source. The emission model-independent polarization angle (PA) and polarization degree (PD) are estimated using the {\sc{pcube}} method whereas {\sc{xspec}} readable count \textit{I} and Stokes \textit{Q} and \textit{U} spectra are created utilizing the algorithms {\sc{pha1}}, {\sc{phaq}} and {\sc{phau}}, respectively, for all units in the $2-8$ keV energy range. We have used the unweighted method and response matrices v012 of {\sc{ixpeobssim}} to produce all the polarization products.

We have rebinned \textit{I}, \textit{Q}, and \textit{U} using {\sc{ftgrouppha}} where the \textit{I} spectrum is grouped using an optimal binning algorithm employing the response file. Subsequently, Stokes spectra are grouped using the \textit{I} spectrum as a template file. We simultaneously fit the count spectra from \textit{NICER} (in the $0.3-8$ keV) and \textit{Swift}/XRT (in the $0.4-5$ keV) and Stokes spectra (\textit{Q}, \textit{U}) from \textit{IXPE} (in the $2-8$ keV) using the X-ray spectral fitting package {\sc{xspec}} version 12.13.1.


\begin{table*}
\centering
 \caption{The polarization degree (PD) and polarization angle (PA) for LMC X-3 using all DUs were extracted using the PCUBE algorithm. The statistical significance of PD within each energy band is given in brackets alongside each PD value. The statistical significance of all the PA values is well above $7 \sigma$.   }
 \begin{center}
\scalebox{0.95}{%
\hspace{-0.5cm}
\begin{tabular}{cccccc}
\hline
Energy (keV) & $2-8$ & $2-2.83 $ & $2.83-4.00 $ & $4.00-5.66 $ & $5.66-8.00 $ \\ \hline
PD (\%) & $2.9\pm{0.4}~(7.3 \sigma)$ & $2.5\pm{0.6}~(4.2 \sigma)$ & $2.8\pm{0.5}~(5.6 \sigma)$ & $3.3\pm{0.9}~(3.7 \sigma)$ & $8.9\pm{2.5}~(3.6 \sigma)$ \\ 
PA (deg) & $135.1\pm{3.8}$ & $136.2\pm{6.7}$ & $130.5\pm{5.5}$ & $141.6\pm{8.1}$ & $134\pm{8}$  \\ 
MDP$_{99}$ (\%) & $1.2$ & $1.7$ & $1.6$ & $2.8$ & $7.4$ \\ 
\hline
\end{tabular}}
\end{center}
\label{pol_table}
\end{table*}

\section{Results}
\label{results}
We carry out the standard model-independent ({\sc{pcube}}) and spectro-polarimetric ({\sc{xspec}}) approach to study the polarization properties of LMC X-3. In the {\sc{pcube}} method, the Stokes parameters (\textit{I}, \textit{Q}, and \textit{U}) are calculated by summing over the observed photoelectric emission angle for each \textit{IXPE} event from which one can recover the polarization angle (PA) and polarization degree (PD) using usual formulae (see equations 21 and 22 in \citealt{kis15}). In the {\sc{xspec}} method, polarization models like {\sc{polconst}}, {\sc{polpow}} or {\sc{pollin}} are used along with basic radiative models to fit the count and Stokes spectra to determine the values of PA and PD.

\subsection{Polarization properties using {\sc{PCUBE}}}
In the full \textit{IXPE} energy range of $2-8$ keV, we estimate a polarization degree of $2.9 \pm 0.4 \%$ ($>7\sigma$ significant) at a polarization angle of $135.1^{\circ} \pm 3.8^{\circ}$ ($>7\sigma$ significant) for the source. We have combined the measurements of all the detectors to improve the confidence interval over PA and PD of the observation. Our results are consistent with \citet{majumder24} and \citet{svoboda24}, which have been reported simultaneously with this work. The polarization is well above the minimum detectable polarization (MDP$_{99}$) of 1.2\% for the source at a 99\% confidence interval. Figure~\ref{2_8_stoke} shows the 1$\sigma$, 2$\sigma$, and 3$\sigma$ confidence contour plots for PA and PD in the $2-8$ keV band. The values of the polarization parameters are listed in Table~\ref{pol_table}. Unless stated, the reported error on each parameter is at the 1$\sigma$ confidence level. 

Further, we extract the polarization properties in four energy bands using the {\sc{pcube}} algorithm and summed the values for each detector for accuracy. It is apparent from Figure~\ref{pa_pd_energy} that the PD is energy-dependent. Upto 5.66 keV the PD remains constant around $3\%$ through the first three energy bands, but increases significantly to $\sim$ 9 \% in the $5.66-8$ keV band. The values of polarization parameters along with their statistical significance and MDP$_{99}$ are listed in Table~\ref{pol_table}. The PA, on the other hand,  is  $\sim 135^{\circ}$ and energy-independent. We also check the time evolution of PA and PD by dividing the data into seven almost equal time bins and running the algorithm for each. As shown in Figure~\ref{pa_pd_time}, we can measure the PA and PD significantly only in two time bins. Thereby, it will require longer exposures to determine any significant variations in the PD and PA with time.

\subsection{Spectro-polarimetric fitting using {\sc{XSPEC}}}
We start by exploring the radiative properties of LMC X-3 by modelling only the \textit{NICER} ($0.3-8$ keV) and \textit{Swift}/XTI ($0.4-5$ keV) spectra simultaneously. We first fit the spectra with the model combination {\sc{constant*tbfeo*(diskbb+powerlaw)}} with abundances from \citet{wi00} and cross-sections from \citet{ve96}. The multiplicative factor {\sc{constant}} accounts for the cross-calibration factor between \textit{NICER} and \textit{Swift}/XTI spectra. The {\sc{tbfeo}} model describes the interstellar absorption and is similar to {\sc{tbabs}} but allows to vary the oxygen and iron abundances of the interstellar absorber. The {\sc{diskbb}} and {\sc{powerlaw}} components describe the thermal disc and non-thermal Compton emissions from the source. The fitting gives a $\chi^2$ of 312 for 276 d.o.f. For \textit{NICER}, we use the systematics as obtained from {\sc{nicerl3-spect}}, whereas, for \textit{Swift}/XRT, we have taken a systematic error of 3\%. Letting the O abundance free, it could be constrained well to a value of $\sim$ 0.55 times the solar abundance. To check if a physical model like {\sc{nthcomp}} can be used to characterize the high energy emission, we fit the data with another model combination, {\sc{constant*tbfeo*(diskbb+nthcomp)}}. This yields a $\chi^2$ of 403 for 277 d.o.f, poorer than the previous case, and thereby we refrain from using this model for further spectro-polarimetric fitting. 

Next, we fit the \textit{I} spectra from \textit{NICER} and \textit{Swift}/XRT and the \textit{Q}, and \textit{U} spectra from \textit{IXPE} simultaneously. We first try the model combination of {\sc{constant*polconst*tbfeo*(diskbb+powerlaw)}} where {\sc{polconst}} is a multiplicative constant, energy-independent polarization model. This gives a $\chi^2$ of 479 for 362 d.o.f. We have not taken any systematic error for \textit{IXPE} spectra, and all the parameters were free during the fitting. The PA and PD values are inconsistent with the {\sc{pcube}} results. This is expected, as we saw in the last section that the polarization degree is energy dependent.

\begin{figure*}
\centering\includegraphics[scale=0.65,angle=-90]{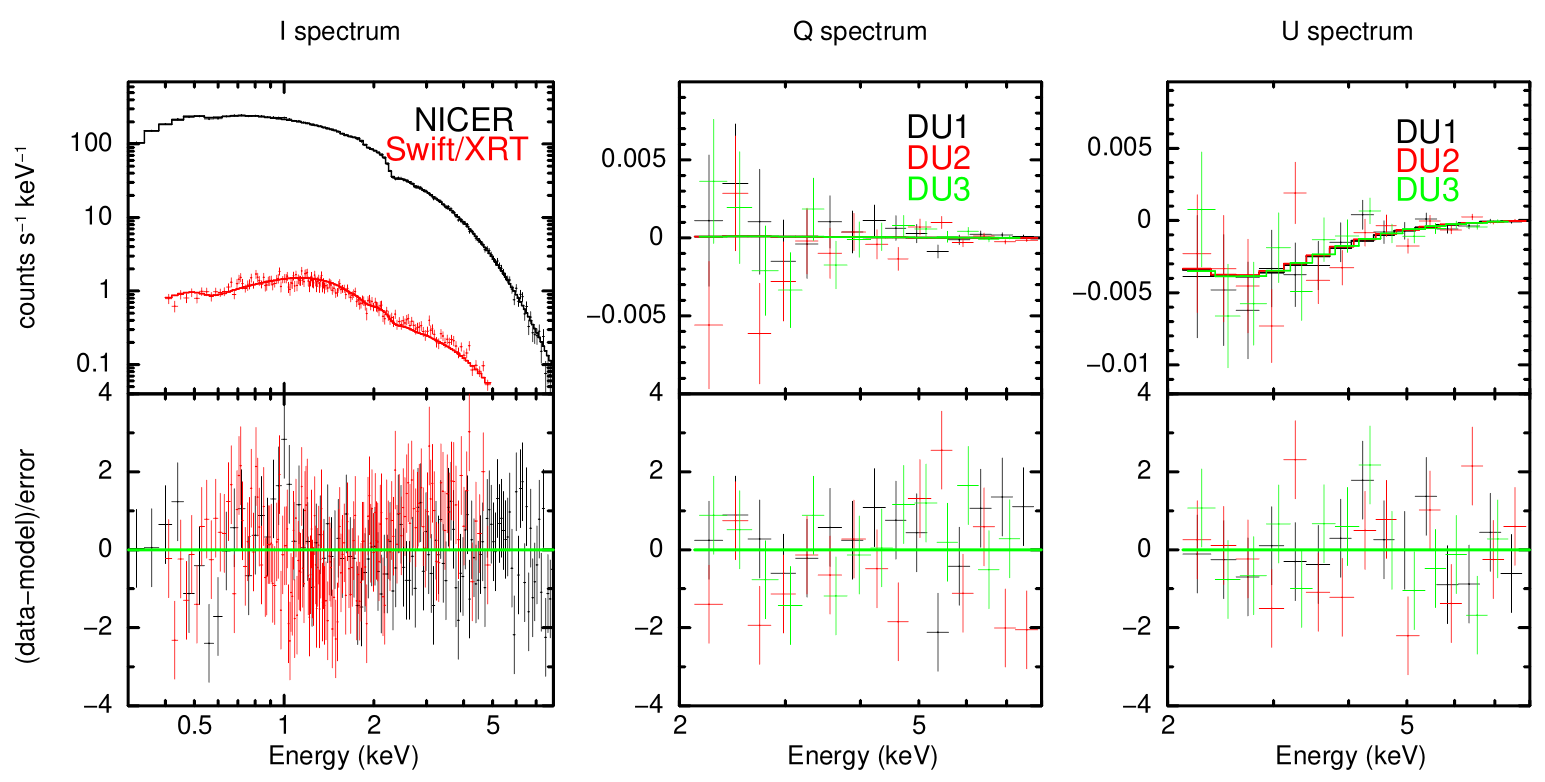}
\caption{The top left panel presents the results of the simultaneous fits to the {\it{NICER}} and {\it{Swift/XRT}} spectra, while the top middle and top right panel shows the {\it IXPE} Stokes spectra, $Q$, and $U$, respectively. The model used for fitting the data is {\sc{polpow*tbfeo*(diskbb+powerlaw)}}. The bottom panels display the residuals of the best-fitting model. For the best-fitting parameters, refer to Table \ref{spectra_table}.}
\label{spectra_I_Q_U}
\end{figure*}

Afterwards, we try the model {\sc{constant*polpow*tbfeo*(diskbb+powerlaw)}} where {\sc{polpow}} is a multiplicative, power-law energy-dependent polarization model. During fitting, we have frozen $\psi_{index}$ to zero because of the energy independence of the polarization angle. The $\chi^2$ for the best-fit comes out to be 412 for 361 d.o.f, better than the previous model. All the best-fit parameters are given in Table~\ref{spectra_table}, and the simultaneously fitted \textit{I}, \textit{Q}, and \textit{U} spectra for all detectors along with residuals are shown in Figure~\ref{spectra_I_Q_U}. The estimated value of PA is $135.7^{\circ} \pm 3.7^{\circ}$, and of PD\footnote{We estimate the PD by integrating the {\sc{polpow}} model in the $2-8$ keV band.} is $5.1 \pm 3.4 \%$\footnote{3$\sigma$ error bar}, which are consistent with {\sc{pcube}} results within the error bars as shown in Figure~\ref{2_8_stoke}.

\begin{table}
\centering
 \caption{{\it{NICER}}, {\it{Swift/XRT}} and \textit{IXPE} best-fitting spectral parameters for LMC~X$-$3 with the model {\sc{constant*polpow*tbfeo*(diskbb+powerlaw)}}.}
 \begin{center}
\scalebox{1.1}{%
\begin{tabular}{lll}
\hline
 Component    & Parameter &  Value\\ \hline
polpow  & $A_{\rm norm}$ ($10^{-3}$) &  $5.4 \pm{3.8}$\\
polpow  & $A_{\rm index}$ &  $-1.4 \pm {0.5}$\\
polpow  & $\psi_{\rm norm}$ (degrees) & $135.7\pm{3.7}$\\
TBfeo     & $N_H$ ($10^{20}$ cm$^{-2}$) &  $5.99\pm{0.46}$ \\
TBfeo     & $O$ &  $0.57\pm{0.15}$ \\
powerlaw   & $\Gamma$  & $2.4\pm 0.1$\\ 
powerlaw   & $norm$ ($10^{-2}$) &  $2.5\pm 0.3$\\
powerlaw & Flux$^{\dagger}$ ($10^{-10}$ erg cm$^{-2}$ s$^{-1}$) & $1.20\pm 0.03$ \\
diskbb  & $T_{in}$ (keV)  & $1.062 \pm {0.008}$\\
diskbb   & norm &  $24.6 \pm {0.64}$ \\
Total Flux$^{\dagger}$ & ($10^{-10}$ erg cm$^{-2}$ s$^{-1}$) & $7.53\pm 0.02$ \\
\hline
$\chi^2$/dof & & 412 / 361 \\
\hline
\multicolumn{3}{l}{$^{\dagger}$ Total unabsorbed flux in the $0.3-8$ keV range.}
\end{tabular}}
\end{center}
\label{spectra_table}
\end{table}

\section{Discussion and Summary}
\label{discuss}
We report the first measurement of X-ray polarization in the soft state of the extragalactic Black hole X-ray binary (BXB) source LMC~X$-$3 using \textit{IXPE} in the $2-8$~keV energy band. The polarization degree (PD) is $\sim$ 3~\% at a polarization angle (PA) of $\sim$ 135~$^{\circ}$. The lightcurves from {\it{IXPE}}, {\it{NICER}} and {\it{Swift/XRT}} show no significant variation of the count rate of LMC~X$-$3 across the $2-8$~keV, $0.2-12$~keV and $0.3-8$~keV bands, respectively (see Figure \ref{lightcurve}). Simultaneous \textit{NICER} and \textit{Swift}/XRT spectra of LMC~X$-$3 exhibit multi-temperature black body emission peaking at around 1 keV. The spectra also display a significant contribution from a thermal Comptonization component with a power law index of approximately 2.4. The PD increases from $\sim$3 \% in the 2--2.83~keV band to $\sim$9 \% in the 5.66--8.00~keV band while the PA shows no dependence on energy (see Figure~\ref{pa_pd_energy}). The energy dependence of the PD is similar to what has been reported for the BXB sources 4U 1630$-$47 \citep{rawat_2023,rawat_2023b, Kushwaha_2023, Ratheesh_2023,rodriguez_2023} and Cygnus X$-$1 \citep{kr22}.  


Employing radiative transfer simulations, \citet{poutanen_2023} computed the dependence of the PD with the inclination angle of the source for both a slab coronal model and a hot inner flow model, accounting for relativistic plasma effects. In the static slab corona model, where the corona covers the cold accretion disk and seed photons originate from it, they found that the PD reaches approximately 4\% for high-inclination sources (i.e., $i\sim70^\circ$) at 4 keV (see top panel of Figure 3 in \citealt{poutanen_2023}). Considering a high inclination angle of LMC X$-$3 \citep{orosz_2014}, our observed polarization fraction of around 3\% (within a 3-sigma range, as shown in Figure \ref{2_8_stoke}) is consistent with this result. Furthermore, we have detected a significant increase in the PD above 5 keV, as illustrated in Figure \ref{pa_pd_energy} and tabulated in Table \ref{pol_table}. \citet{poutanen_2023} also found an increase in the PD with energy for the case of the static corona for two different inclination angles, $i=30^\circ$ and $i=60^\circ$, and a specific set of disc and coronal parameters, $KT_{bb}=0.1~keV$, $KT_e=100~keV$ and $\Gamma = 1.6$. They show that the PD rises from $\sim$2 \% to $\sim$5 \% for $i=60^\circ$ in the IXPE band of 2--8 keV. Our observations indicate an increase of the PD of up to around 9\%, with a lower limit of 6.5\% within the same energy range, which is closer to what \citet{poutanen_2023} found.  The minor difference between their results and ours could be due to the spectral parameters we estimated, which could change the number of scatterings at higher energies and consequently influence the PD value \citep{poutanen_2023}. Moreover, the slab coronal model is known to generate excessively soft X-ray spectra \citep{stern95}, and LMC X-3 also possesses an unusually soft X-ray spectrum comprising an ultrasoft and a high-energy component during its soft state \citep{white_1984}. So, the spectral analysis also gives a hint for slab-like corona geometry, but observed polarisation measurements along with the simulation results provide additional support.

\citet{Schnittman_2010} also calculated X-ray polarization for accreting black holes for three different coronal geometries, and they observed a consistent trend of increasing PD and a transition from horizontal to vertical polarization across all the geometries. However, they also found that in the spectral state characterized by an inner disk temperature of 1 keV and an electron temperature of 100 keV, the level of polarization is $\sim$3 \% for sandwich geometry but lower for the clumpy and spherical coronae at high inclination angles. It should be noted that their sandwich geometry is somewhat like the slab-like corona, which we argue is a likely geometry to explain the polarization of LMC X-3.\\

\textbf{\textit{}}
\citet{Poutanen_1996} highlighted that the reversal of the PD sign is a distinctive characteristic associated with the slab corona geometry. Here, the positive polarization indicates that the electric field vector is predominantly perpendicular to the slab corona, while negative polarization indicates that the electric field vector is predominantly parallel to the slab corona. Using the radiative transfer code of \citet{poutanen_2023} and considering a slab coronal geometry for seed photons temperature of kT$_{bb}$ = 0.8 keV and Comptonizing medium of temperature kT$_{e}$ = 10 keV, \citet{Podgorny_2023_LMX_X1} found that the PD changes sign from negative to positive at $\sim$ 5 keV (see Figure 8 of  \citealt{Podgorny_2023_LMX_X1}). They suggest that the overall polarization arises from emission polarized in both the parallel and perpendicular directions, which causes a low PD in the system. However, in this work, we found that the polarization fraction remains at $\sim$3 \% in the 2--5.66 keV band and increases to $\sim$9 \% in the 5.66-8 keV band as quoted in Table \ref{pol_table}. But, it is to be noted that the temperature of the seed photons, the Comptonizing medium, and the geometry of the coronal medium significantly influence the energy at which the sign reversal of the PD occurs here and thereby could be the cause of the absence of any sign reversal in our observations.

In the context of BXBs, assuming a static hot inner flow geometry, the anticipated polarization level is approximately 7-8\% at 4 keV, as demonstrated in the middle and bottom panels of 
Figure 3 in \citet{poutanen_2023}. Furthermore, the polarization fraction can be enhanced when accounting for relativistic plasma. \citet{poutanen_2023} has shown that for Cygnus X$-$1, the polarimetric results cannot be explained through static coronal models, and thus an out-flowing plasma with a mildly relativistic velocity is required. \citet{dexter_2023} has also proposed bulk Comptonization of coronal emission in a mildly relativistic wind or jet for Cygnus X$-$1 to explain the observed polarization fraction. Unlike Cygnus X$-$1, the static slab corona geometry provides a successful explanation for the polarimetry findings in the case of LMC~X$-$3. It is worth noting that the possibility of an enhancement of the PD due to disk winds, as reported by \citet{kosenkov_2020} for the black-hole binary MAXI J1820+070, or PD suppression due to disk winds, as suggested by \citet{Tomaru_2023} for 4U 1630$-$47, cannot be ruled out. \citet{majumder24} proposed that a direct and reflection emission in the partially ionized atmosphere \citep[as suggested by][for the black hole source 4U 1630+47]{Ratheesh_2023} could be the cause of the observed energy-dependent polarization. However, in the spectra of LMC X$-$3, we have not detected any signature of relativistic disk winds which can show the presence of any ionized atmosphere.

 If disk self-irradiation produces polarization, the PA should align with the jet axis, i.e. perpendicular to the disk plane in BXBs \citep{sc09}. Studies on Cygnus X$-$1 conducted by \citet{ch18} and \citet{kr22} utilizing {\it{PoGO+}} and {\it IXPE} observations, respectively, have confirmed the alignment of the polarization angle with the radio jet. Similarly, optical polarimetry conducted by \citet{kosenkov_2020} has indicated that the polarization angle aligns with the jet axis for MAXI J1820+070 in the rising hard state. Unfortunately, for LMC~X$-$3, the jet angle is unknown as no radio counterpart has been reported \citep{fender_2000}, so we cannot make any comparison.

\section*{Acknowledgements}
We are grateful to an anonymous reviewer for their constructive comments, which helped us significantly improve the quality of the manuscript.
This work used data from the UK Swift Science Data Centre at the University of Leicester.
This research has used data from the High Energy Astrophysics Science Archive Research Center Online Service, provided by the NASA/Goddard Space Flight Center. DR acknowledges 
Centre National de la Recherche Scientifique (CNRS) for financial support.
MM acknowledges support from the research program Athena with project number 184.034.002, which is (partly) financed by the Dutch Research Council (NWO). 
\section*{Data Availability}
The \textit{NICER}/XTI, \textit{Swift}/XRT, and \textit{IXPE} observations used in this work are available at the \href{https://heasarc.gsfc.nasa.gov/db-perl//W3Browse/w3browse.pl}{HEASARC website}.

\bibliographystyle{mnras}
\bibliography{manuscript}






\bsp	
\label{lastpage}
\end{document}